\documentclass[twocolumn,showpacs,preprintnumbers,amsmath,amssymb,prb,a4]{revtex4}

\usepackage{graphicx}

\newcommand{\Na}{NaV$_2$O$_5$}
\newcommand{\Ca}{CaV$_2$O$_5$}

\newcommand{\fs}{\footnotesize}

\begin{document}

\title{Optical properties, electron-phonon coupling,  and Raman scattering of vanadium ladder compounds}
\author{J.~Spitaler$^1$}
\email{juergen.spitaler@uni-graz.at}
\author{E. Ya. Sherman$^{1,2}$}
\author{H. G. Evertz$^2$}
\author{C. Ambrosch-Draxl$^1$}
\affiliation{$^1$Institut f\"{u}r Theoretische Physik, University Graz, Universit\"{a}tsplatz
5, A-8010 Graz, Austria}
\affiliation{$^2$Institut f\"{u}r Theoretische Physik, Technical University Graz, Petersgasse 16, A-8010 Graz, Austria}
\date{\today}
\begin{abstract}
The electronic structure of two V-based ladder compounds,  the quarter-filled  \Na\ in the symmetric phase 
 and the iso-structural half-filled \Ca, is investigated by
{\em ab initio} calculations. Based on the bandstructure 
we determine the dielectric tensor $\varepsilon(\omega)$ 
of these systems in a wide energy range. 
The frequencies and eigenvectors of the fully symmetric A$_{g}$ phonon modes  and the 
corresponding electron-phonon and spin-phonon coupling parameters are also calculated 
 from first-principles. 
We determine the Raman scattering intensities of the A$_g$ phonon modes as a function 
of polarization and frequency of the exciting light.
 All results, i.e.~shape and magnitude of the dielectric function, phonon frequencies 
 and Raman intensities show very good agreement with 
available experimental data.
 
\end{abstract}

\pacs{71.15.Mb, 63.20.Kr, 78.30.-j, 71.27.+a}

\maketitle

\section{Introduction}\label{intro}

\begin{figure}[h*]
\begin{center}
\includegraphics[width=0.44\textwidth]{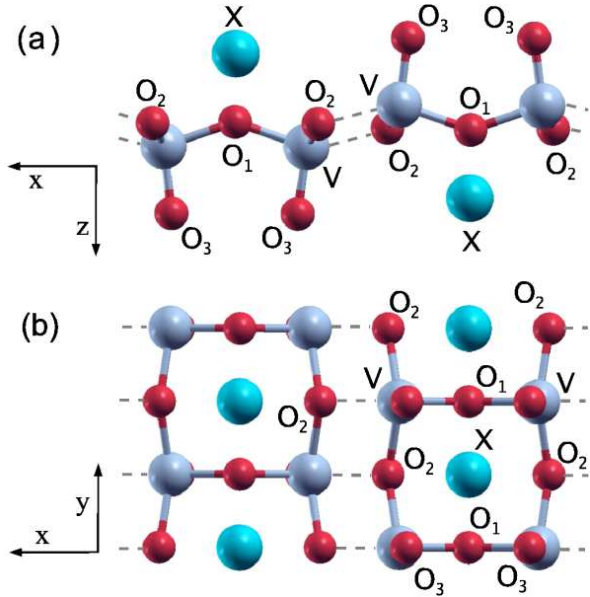}
\end{center}
\vspace{-4mm}
\caption{ The crystal structure of \Na\ and \Ca. (a)~the view along the ladders,
          (b) the view from the top. The dashed lines correspond to relatively 
          weak bonds between the ladders. X indicates either Na or Ca, respectively.}
\label{fig:structure}
\end{figure}

The vanadium-based ladder compounds \Na\ and \Ca\ are interesting examples of systems where charge, spin, and lattice degrees of freedom are coupled to each other. Like materials with magnetically active Cu ions forming ladder-like structures,\cite{Dagotto96,Lemmens99} they show unusual physical properties due to a strong interaction  of all degrees of freedom. 
    The main building block of their unit cells is a ladder formed by V-O rungs and V-O legs, as it is shown in Fig.~\ref{fig:structure}. 
	Both materials crystallize in the orthorhombic space group  $Pmmn$ ($D_{2h}^{13}$) with two formula units per unit cell.

Electrons in these systems can move within a rung, between rungs within a 
ladder, and between different ladders. 
The upper occupied electron states are formed by $d_{xy}$ electrons of V with 
a slight admixture of oxygen $2p$ orbitals.  
In \Na, one $d_{xy}$ electron of V is shared by two sites within a rung 
which leads to a quarter-filled character\cite{Smolinski98} and makes this compound a dielectric, 
since the hopping of an electron between two rungs, which would produce a doubly occupied and an empty rung,
requires more energy than can be provided by the hopping matrix element along the ladders. 
At a critical temperature $T_c$ close to 35 K, \Na\ undergoes a transition  to a spin-gapped state, as 
first reported by Isobe and Ueda in Ref.~\onlinecite{Isobe96} based on the analysis of magnetic
susceptibility measurements. Evidence for such a transition was also obtained in Raman scattering experiments.\cite{Weiden97} 
This behavior, in some aspects analogous to the spin-Peierls transition
observed in the inorganic chain compound CuGeO$_3$  at $T_{SP}=13.5$~K, \cite{Hase93}
is accompanied by a disproportion of the V charges ($4.5\pm\delta$), 
a relatively large lattice distortion, and the formation of an ordered charge pattern. At $T\ll T_c$ the ions are displaced from their
equilibrium positions in the high-temperature phase 
by distances of the order of 0.05 \AA.  These displacements 
which give evidence for a strong electron-lattice coupling have been observed in
X-ray diffraction\cite{Luedecke99,Smaalen02,Nakao00} and can be estimated from infrared 
\cite{Popova02} and Raman scattering experiments.\cite{Fischer99}
They stabilize the zig-zag ordered phase,\cite{Seo98,Mostovoy00}
being probably the crucial element that determines the physics of the charge ordering in this compound.   
At the same time, the coupling to dynamical phonons 
induces strong charge fluctuations near $T_c$ which alter the spin-spin exchange $J$.\cite{Sherman99}
Also the ordering is not completely static, even at low temperatures, since it is influenced by lattice vibrations, as was found in electron spin-resonance experiments.\cite{Nojiri00}
At the same time, one could expect that charge ordering 
accompanied by a lattice distortion will show up, for example,  
in the dielectric function with decreasing temperature. 
However, a comparison of optical spectra taken at the low-
and high-temperature phase, respectively, exhibits only small differences below $T_c$,\cite{Presura00} while Raman spectra reveal large changes in the electronic background and 
show many new peaks which could have either magnetic or phononic origin when going below $T_c$.\cite{Fischer99} 
New peaks are also clearly seen in the low-temperature infrared spectra.\cite{Popova02} 
A full understanding of these new modes is still lacking. 

In \Ca, each V ion has spin $\frac{1}{2}$ and it can therefore be described by a generalized Heisenberg model of the spin-spin interaction.
In this case, phonons directly influence the exchange constants, 
and, therefore, lead to a modification of the magnetic properties.  
The spin gap in this compound, arising due to different exchange along the 
legs and along the rungs, is of the order of 500~K (0.05~eV).\cite{Iwase86} 
We mention that the Heisenberg spin-spin exchange parameters and, in turn, 
the strength of spin-phonon coupling  
depend on the electron on-site energies and the hopping matrix elements which 
form the bandstructure and influence the dielectric function.
For this reason, the experimental and theoretical investigation 
of the optical conductivity, 
the lattice dynamics, and the electron-phonon and spin-phonon coupling 
can provide a clue to the properties of \Na\ and \Ca\ and shed light on the 
origin of the phase transition in the former. 

\Na\ and \Ca\ have been the subject of intensive 
theoretical investigations, both by first principles based on density functional theory (DFT) 
and by model calculations.\cite{Horsch98,Yushankhai01,Hubsch01,Aichhorn02,Mostovoy02,Vojta01} 
In first-principles calculations, the tight-binding linear muffin-tin orbitals 
(LMTO) method\cite{Popovic99,Yaresko00}  in the atomic sphere approximation  (ASA) 
and the linear combination of atomic orbitals \cite{Wu99} (LCAO) approach, which directly included the Hubbard repulsion on the V sites, were applied to investigate the bandstructure of \Na. 
More recently, Mazurenko {\it et al.} \cite{Mazurenko02} combined DFT calculations
with dynamical mean field theory aiming at understanding the insulating behavior 
of this compound in the high-temperature phase. The tight-binding 
LMTO method was also applied to \Ca\ to obtain the spin exchange  and effective 
four-band tight-binding model parameters for this compound.\cite{Korotin00}
In addition, approaches based on quantum chemistry \cite{Bernert00,Suaud00,Suaud02,Hozoi02} 
have been applied to \Na.  An interesting feature of the approaches used in Refs.~\onlinecite{Suaud02} 
and \onlinecite{Hozoi02} is that being based on the strong coupling of V and O1-$p_y$ orbitals,
they attribute the phase transition to the ordering of the spin rather than the charge subsystem. 

While the bandstructures of \Na\ and \Ca\ are rather well understood on the 
first-principles basis, the analysis of their optical properties, 
lattice dynamics, electron-phonon coupling and Raman scattering still rely 
on various model assumptions. Having been well investigated experimentally,
these properties require a detailed theoretical treatment which does not depend 
on such assumptions. Moreover, the outcome of DFT calculations is further used 
as realistic input parameters for model calculations such as Quantum Monte Carlo 
or Exact Diagonalization techniques.\cite{Aichhorn03,Gabriel03}

As far as \Na\ is concerned, in this paper we concentrate on the high-temperature phase where V ions in the rungs are equivalent.
This enables us to understand its main properties and provides a starting point for the investigation of the low-temperature  phase. 
Since in \Ca\ no structural phase transition is observed, our treatment there holds at any temperature.   
The paper is organized as follows: 
In Section II, we describe the method of calculation and present the
bandstructure for \Na\ and \Ca\ and related results  like the density of states 
and the charge density. We provide the calculated dielectric 
tensor components $\varepsilon_{ii}(\omega)$ ($i=x,y,z$) for these two compounds in Section III.
Section IV includes theoretical phonon frequencies and eigenvectors, as well as 
electron-phonon and spin-phonon coupling parameters. The phonon-induced changes 
in the dielectric function and the corresponding phonon Raman spectra will be presented in 
Section V. Finally, a summary of the results and suggestions for further investigations are given in the Conclusions.

\section{Calculations of the electronic structure}

\subsection{Computational Methods}
All  bandstructure calculations are performed within density functional theory (DFT) 
using the full-potential augmented planewaves + local orbitals (FP-APW+lo) \cite{Sjostedt00} formalism implemented in the WIEN2k code.\cite{wien2k} 
Exchange and correlation terms  are described within the generalized gradient approximation (GGA).\cite{Perdew96} The atomic sphere radii are chosen as
1.6~a.u.~for V, 1.4~a.u.~for the O atoms and Na, and 1.5~a.u.~for Ca. In both compounds, all atomic positions have been relaxed starting from the experimentally measured ones as given in Ref.~\onlinecite{Onoda96} for \Ca\ 
and Ref.~\onlinecite{Smolinski98} for \Na.
In \Na, for example, the shifts of the ions due to the relaxation of the structure are up to approximately 0.015 \AA\  (for O1 and O2 in $z$ direction) with an energy gain of roughly 30~meV per unit cell.

Comparing the two materials, they have slightly different lattice constants and ion coordinates.
For example, since V is less positively charged in \Ca, the lengths of the V-O bonds are slightly larger than those in \Na. Specifically, the optimized values for the V-O3 and V-O1 bond lengths are 1.62 (1.67) \AA\ and 1.82 (1.85) \AA\ in \Na\  (\Ca). At the same time, the spacing between Ca and the O1 plane (2.39 \AA) is  smaller than the Na-O1 plane distance in \Na\  (2.44 \AA), since the bigger Ca ion exhibits a stronger Coulomb interaction with oxygen compared to Na. The geometry relaxation allows to make the calculations not directly relying on the 
experimentally measured structural data and thereby leads to small quantitative 
differences compared to the \Na\ calculations performed by Smolinski {\it et al.}\cite{Smolinski98} 
At the same time, this provides the energy scale related to unit cell distortions as it is realized for example in the low-temperature phase.

We do not include correlational effects by using an LDA+$U$ approach in our 
calculations, but we have estimated the Hubbard $U$ for \Na\ and \Ca\ from our data by the following procedure.
Similarly to what is described in Ref.~\onlinecite{Smolinski98},
we have added a small amount of electronic charge to the system and 
estimated $U$ from the resulting change of the V bands. Charge neutrality 
was accounted for by two different procedures: When putting the positive charge 
on the Na sites, the resulting $U$ was estimated to be 2.8~eV for \Na\  
(averaged over the Brillouin zone (BZ)) in good agreement with Smolinski {\it et al.}\cite{Smolinski98} 
We preferred, however, to provide the positive charge in terms of 
a uniform background, which leads to a $U$ value of  2.45 eV for both compounds, 
demonstrating that $U$ only weakly depends on the ion's surrounding. 
This procedure has the advantage, that the band energy shifts are much more 
uniform with respect to different $k$ points of the BZ than when the 
additional positive charge is located at the Na sites (differences of 
hundredths of an eV in the former case compared to tenths of an eV in the latter case). 

\subsection{Bandstructure and density of states}

\begin{figure}[h!]
\begin{center}
\includegraphics[width=0.45\textwidth]{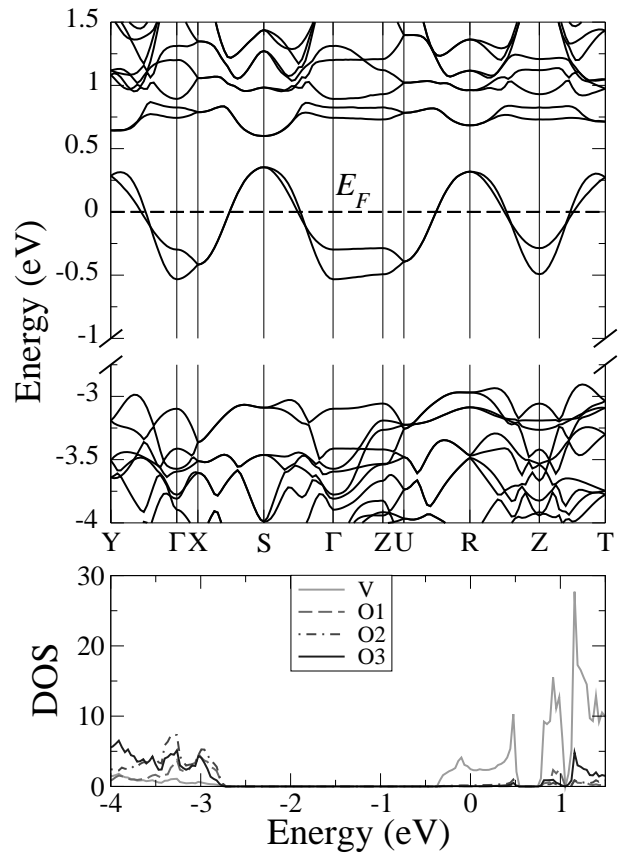}

\end{center}
\caption{ Bandstructure and density of states (in states per unit cell and eV) of \Na.}
\label{fig:bs_na}
\end{figure}

\begin{figure}[h!]
\begin{center}
\includegraphics[width=0.45\textwidth]{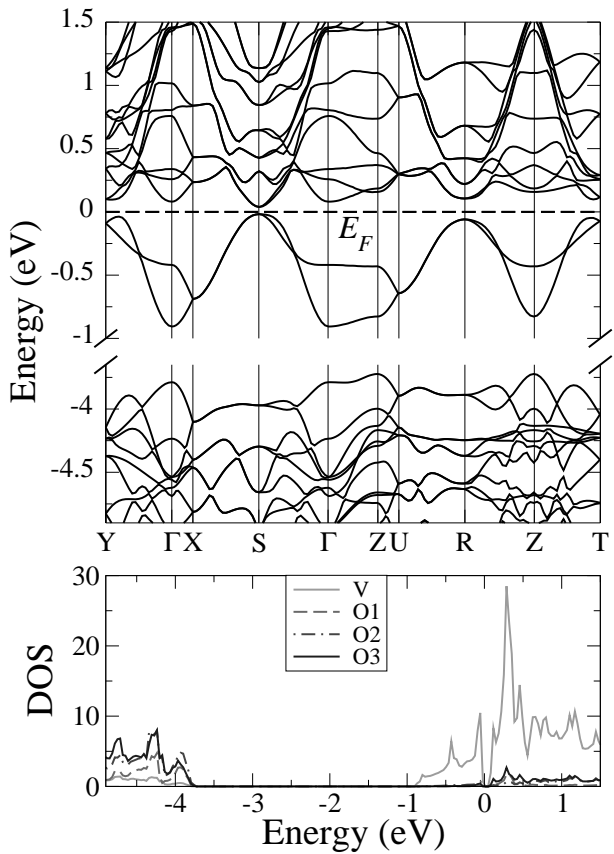}
\end{center}
\caption{Bandstructure and density of states (in states per unit cell and eV) of \Ca.}
\label{fig:bs_ca}
\end{figure}

The bandstructures and densities of states of \Na\ and \Ca\ are shown 
in Figs.~\ref{fig:bs_na} and \ref{fig:bs_ca}, respectively. 
Focusing on \Na\ first, the bands 3-4~eV below the Fermi level are due to O-$2p$ states.
Their smallest distance to the valence band minimum (at the $\Gamma$ point) 
will be denoted as $E_g$ later in the text. The unoccupied states 
exhibit mainly V character with a small admixture of oxygen. 
The bands intersecting the Fermi level in $k_y$ direction are formed  by the bonding combination of V orbitals. 
Their  dispersion is due to hopping along the ladders while the splitting 
of this pair of bands at the $\Gamma$ point is due to inter-ladder hopping. 
The two bands just above the Fermi level originate from the antibonding 
combination of V-$d_{xy}$ states.  
The situation for \Ca\ is similar, where some quantitative differences will be discussed below  and in the context of its optical properties.

The bands can be mapped onto a tight-binding model with the one-ladder 
parameters $t_{\perp}$ of the in-rung hopping, and $t_{\parallel}$ representing the hopping along the ladder.
The theoretically determined values for \Na\ are $t_{\perp}=0.387$~eV and $t_{\parallel}=0.175$~eV, 
which is close to the data of Ref.~\onlinecite{Smolinski98}. 
For \Ca, we obtained in the same way  $t_{\perp}=0.321$~eV and $t_{\parallel}=0.143$~eV, 
in agreement with the results of Korotin {\it et al.}\cite{Korotin00} 
who applied the LDA+$U$ technique in their calculations. 
Compared to their results, our hopping matrix elements are slightly increased, which is due to the changes in the interatomic distances as a result of the structural relaxation.
We note that both $t_{\perp}$ and  $t_{\parallel}$, are smaller for \Ca\ than 
for \Na. At the same time, the splitting of the {\it bonding} bands 
in the $\Gamma$ point arising from the inter-ladder hopping is much larger 
in \Ca\  (0.49 eV) than in \Na\  (0.23 eV) since the distance between the 
ladders is smaller in the former. The corresponding inter-ladder hopping 
matrix elements between the closest V atoms of neighboring ladders, $t_i$, are 0.13 and 0.24~eV for \Na\ and \Ca, respectively.    
When the lattice is deformed by a displacement of ions corresponding to a 
phonon mode, the tight-binding parameters as well as the on-site energies change.  
This kind of electron-phonon coupling will be discussed below. 

 The influence of the Hubbard term $U$ on the properties of V-based ladder compounds is widely 
discussed in the literature. It is important to mention that the enhanced electron correlation when accouted for by the Hubbard parameter reproduces the semiconducting behavior with the charge gap close to $2|t_{\perp}|$ .\cite{Horsch98}
Correspondingly, the dispersion along the $y$-axis of the band derived from the bonding combination 
of V-$d_{xy}$ orbitals will be $\pi$ rather than $2\pi$ periodic.\cite{Kobayashi98,Damascelli04}
At the same time, we shall see below that the physical properties determined by the electron density, 
are not strongly influenced by the Hubbard repulsion and can be described reliably within DFT.

\begin{figure}[htb]
\begin{center}
\includegraphics[width=0.5\textwidth]{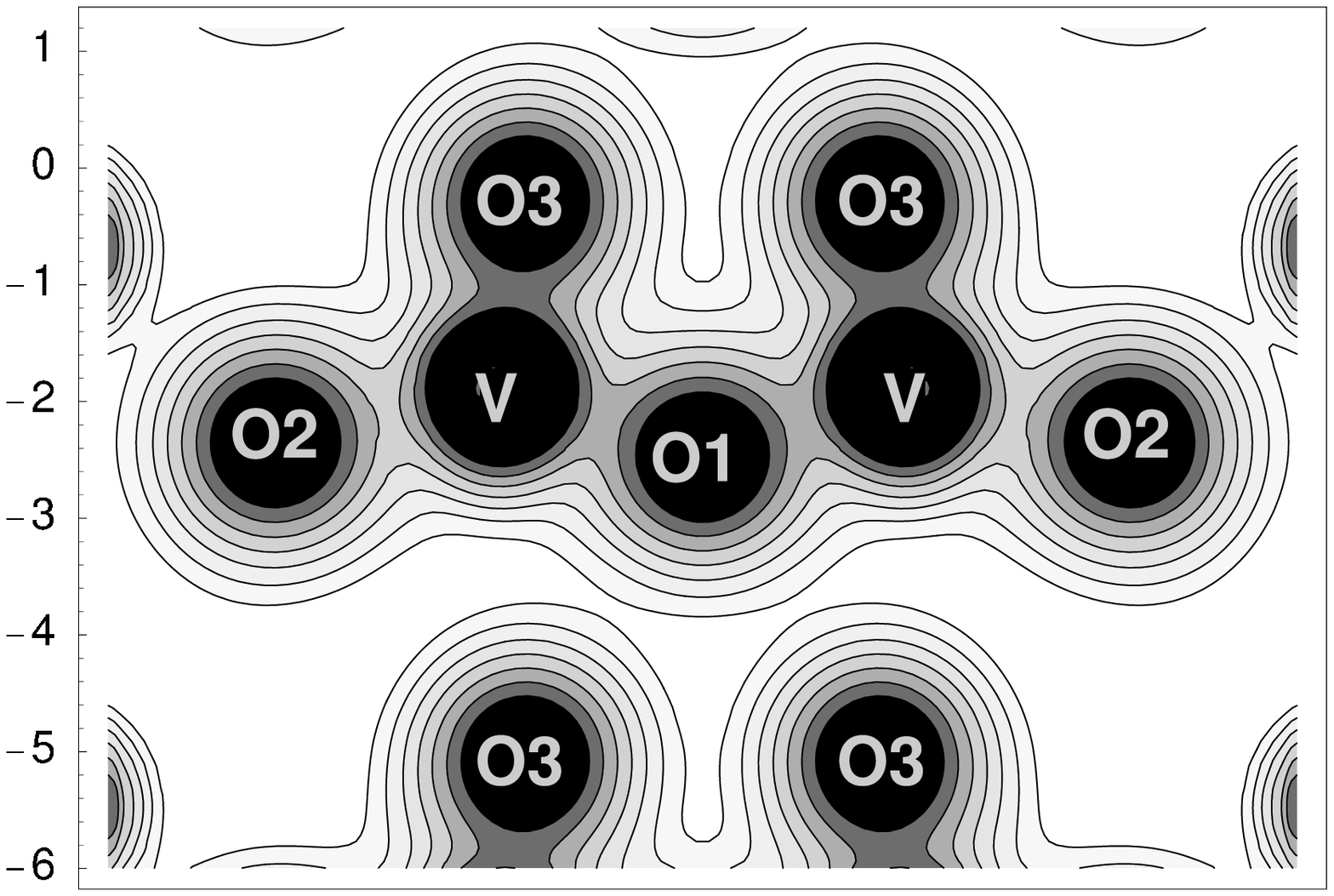}\\
\vspace{-3.5mm}
\includegraphics[width=0.5\textwidth]{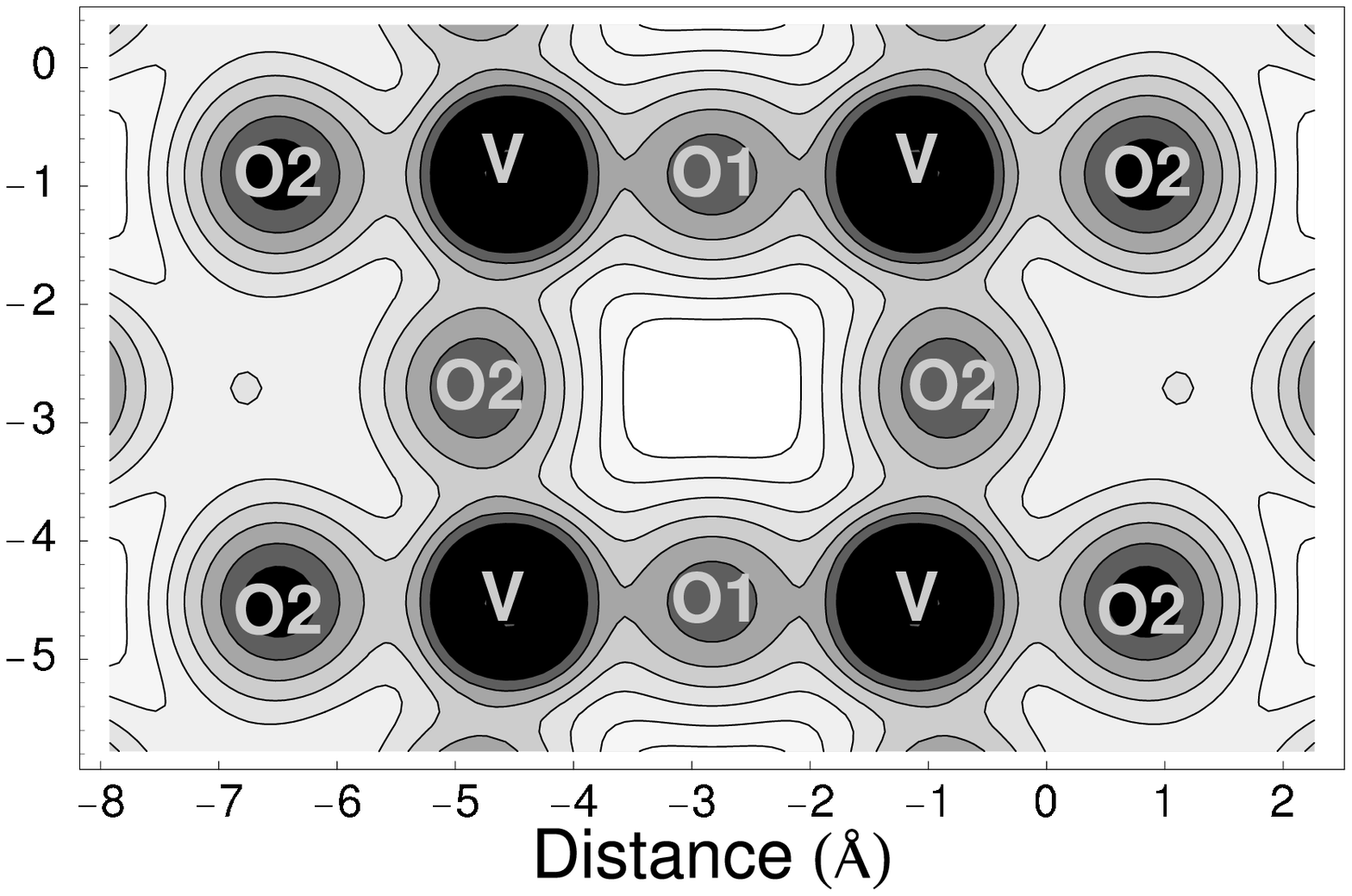}
\end{center}
\vspace{-9mm}
\caption{Valence electron density of \Na\ in the $(x,z)$ plane (upper panel) and the $(x,y)$ plane (lower panel), which both contain the V positions. 
The logarithmic contour lines range from $0.07$ $e$/\AA$^3$
to $2.3$ $e$/\AA$^3$.}
\label{fig:e_dens}
\end{figure}

To illustrate the charge density distribution within the unit
cell, we exemplarily present 
the electron charge density for \Na\ in Fig.~\ref{fig:e_dens} in two perpendicular planes.
The upper panel clearly shows the role of the unit cell asymmetry
on the charge density arising due to the presence of the apical oxygen O3.  
This asymmetry, on the one hand, leads to a strong Holstein-like electron-phonon
coupling, and, on the other hand, diminishes the overlap of the V orbitals with O1 and O2 states,
thus decreasing the hopping matrix elements and correspondingly the 
components of the dielectric tensor.  
In addition, the lower panel shows the preferred orientation of the in-ladder
oxygen states and a relatively small overlap of the orbitals from different ladders.

\section{Dielectric function}

\begin{figure}[htb]
\begin{center}
\includegraphics[width=0.45\textwidth]{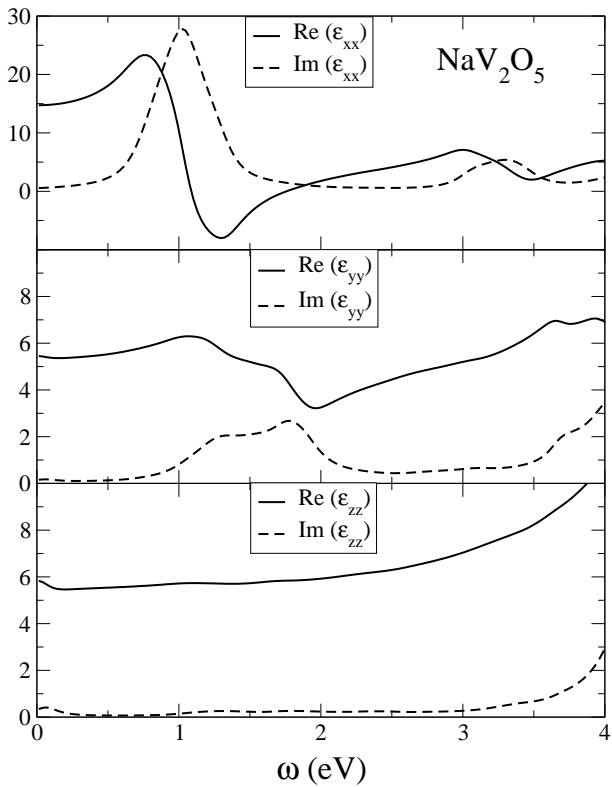}
\end{center}
\vspace{-3mm}
\caption{Diagonal components of the dielectric function $\varepsilon_{ii}(\omega)$ of \Na. }
\label{fig:epsilon_na}
\end{figure}

\begin{figure}[htb]
\begin{center}
\includegraphics[width=0.45\textwidth]{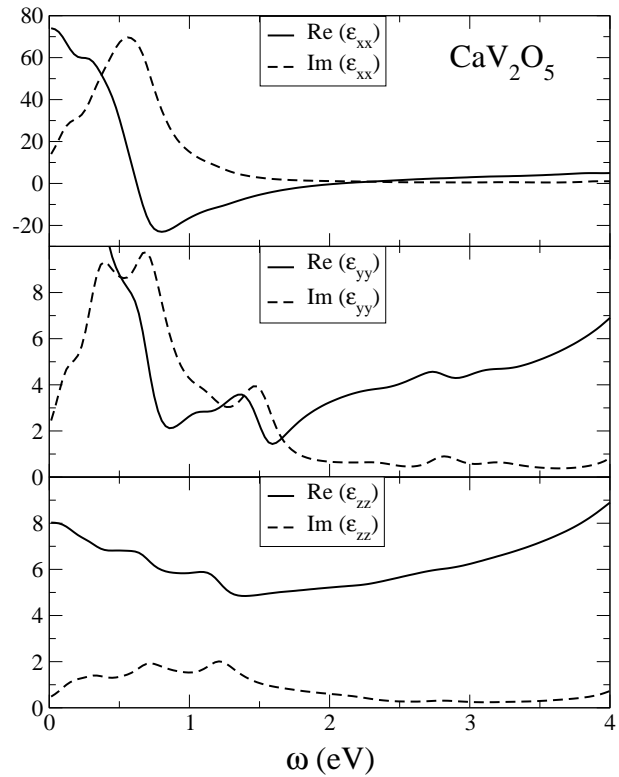}
\end{center}
\vspace{-3mm}
\caption{Diagonal components of the dielectric function $\varepsilon_{ii}(\omega)$  of \Ca.  }
\label{fig:epsilon_ca}
\end{figure}

Figures \ref{fig:epsilon_na}  and  \ref{fig:epsilon_ca}  present the real and imaginary parts 
of the diagonal dielectric tensor components, 
${\rm Re}~\varepsilon_{ii}(\omega)$ and ${\rm Im}~\varepsilon_{ii}(\omega)$, 
in the experimentally measured range, 
where the Cartesian index $i$ corresponds to the light polarization directions. 
${\rm Im}~\varepsilon_{ii}(\omega)$  was calculated within the Random Phase Approximation 
(RPA), based on the Kohn-Sham orbitals including a life-time broadening of the optical inter-band 
transitions of $0.1$ eV, 
while ${\rm Re}~\varepsilon_{ii}(\omega)$  is obtained by Kramers-Kronig transformation. 

Let us first discuss the in-plane response of \Na.  
The most interesting feature of the $xx$ component (light polarized along the rungs)
 is a strong peak at 1.03~eV in agreement with experiment.\cite{Presura00,Golubchik97,Loosdrecht}
An analysis of the interband momentum matrix elements at different electron
 wavevectors $k_y$ shows that the first peak in the $xx$ response arises due 
 to transitions between the bonding and antibonding band states within one V rung. 
 The energy of the peak is larger than 2$t_{\perp}$ because of the band dispersion 
 along the $y$ axis and can be estimated to be  2$(t_{\perp}+t_{\parallel})$, which is 1.12~eV.
 Since the transitions between $d_{xy}$ orbitals have very small matrix elements 
 due to the large V-V distance, the admixture of O1-$p_y$ states in the antibonding state  
 is responsible  for a sizable intrarung transition matrix element. We mention here
 that this admixture provides some support to the arguments of 
 Refs.~\onlinecite{Suaud02} and \onlinecite{Hozoi02}. 
 It rapidly decreases, however, with the increase of $k_y$ due to the 
 corresponding decrease of the oxygen contribution, which was first 
 noticed in Ref.~\onlinecite{Smolinski98}. For this reason the peak mostly  originates
from transitions in the vicinity of the $\Gamma$ point. 
   
The $yy$ component (light polarized along the legs) is dominated by a double-peak structure at 1.27~eV and 1.78~eV, respectively. 
It is considerably weaker than the $xx$ response.  The analysis of the bandstructure reveals that the shoulder at 1.3~eV comes from in-rung transitions. These can contribute to $\varepsilon_{yy}(\omega)$ since at finite $k_x$ values the in-rung states are neither odd nor even with respect to the $x\rightarrow -x$ transformation, and, therefore, can couple to light polarized along the $y$-axis.
The broad maximum  at $\omega=1.8$~eV is related to transitions from the 
bonding V-$d_{xy}$ states to O2-$p_x$ states admixed to V-$d_{xz}$-derived orbitals at approximately 1.7~eV above the Fermi level. 

In both polarizations, the peaks occurring at higher energies ($\omega>3$~eV) originate 
from transitions between O-$2p$ and V orbitals.
For example, the broad feature around 3 eV in the $xx$ spectra arises from transitions
between O1-$2p_z$ states at -3.3 eV and bonding V-$d_{xy}$ states around $E_F$. 
Comparing to experiment, we want to point out that all theoretically obtained features  
reproduce the corresponding experimental findings very well.\cite{Presura00}

In \Ca, the character of the transitions is, in general, the same as in \Na. 
The fact that Ca provides one more valence electron to the system compared to Na has 
two obvious effects: First, the optical response is stronger compared to \Na. 
Second, the peak present in \Na\ around 3~eV is missing, since the bonding V-$d_{xy}$ states 
are occupied and thus don't provide final states for the transitions. 
At the same time, since the in-ladder hopping matrix elements are smaller in \Ca\ than in \Na, 
the spectrum of unoccupied band states in \Ca\ is denser, 
as can be seen in Figs.~\ref{fig:bs_na} and ~\ref{fig:bs_ca}.
For this reason, the low-energy part of the dielectric function ($\omega<1$ eV) of \Ca\ 
shows a more complicated $\omega$ dependence than that for \Na\ since more interband 
transitions are allowed at this spectral range. 
A comparison with experiment for this compound 
which can also provide an experimental test of the applicability of the DFT for the description 
of optical properties of half-filled ladder compounds 
is not possible at present since to the
best of our knowledge no measured data of the dielectric function of \Ca\ single crystals 
are available.

\section{Lattice dynamics}

\subsection{Phonon modes}

\begin{table*}
\begin{tabular}{|cc@{\hspace{3mm}}c|cccccccc|c@{\hspace{2mm}}c|}
\hline
\multicolumn{3}{|c|}{Frequency (cm$^{-1}$)}  & \multicolumn{8}{c|}{Eigenvector} & \multicolumn{2}{|c|}{Assignment} \\  
\hline
\multicolumn{2}{|c}{ Experiment}
 & Theory  &&&&&&&&&&\\[-.7mm]
\fs Ref.~\onlinecite{Fischer99}&\fs Ref.~\onlinecite{Popovic99ssc} & & \fs  V$_x$&  \fs V$_z$& \fs Na$_z$& \fs O1$_z$& \fs O2$_x$& \fs O2$_z$&\fs O3$_x$& \fs O3$_z$ & Refs.~\onlinecite{Popovic99ssc, Popovic02} & This work \\
	\hline
970 & 969 & 996 & \fs  0.04  & \fs  0.25  & \fs -0.01  & \fs  0.01   & \fs  0.01 &  \fs -0.00 & \fs -0.05 & \fs -0.43 & V-O3 stretching   & V-O3 stretching     \\
530 & 534 & 512 & \fs  0.16  & \fs -0.07 &  \fs  0.01  & \fs  0.16   & \fs  0.45 &  \fs  0.06 & \fs -0.05 & \fs -0.01 & V-O2 stretching   & V-O2 stretching     \\
450 & 448 & 467 & \fs  0.42  & \fs -0.06 &  \fs  0.01  & \fs  0.15   & \fs -0.21 &  \fs -0.05 & \fs -0.11 & \fs  0.01 & V-O1-V bending    & V-O1-V bending      \\
422 & 423 & 414 & \fs -0.19  & \fs -0.17  & \fs  0.00  & \fs  0.44   & \fs -0.08 &  \fs -0.18 & \fs  0.18 & \fs -0.14 & O3-V-O2 bending   & O1$_z$+O3-V-O2 bend.\\
304 & 304 & 308 & \fs  0.02  & \fs  0.19  & \fs -0.04  & \fs  0.18   & \fs -0.07 &  \fs  0.39 & \fs  0.19 & \fs  0.09 & O3-V-O2 bending   & O3-V-O2 bending     \\
230 & 233 & 232 & \fs  0.17  & \fs  0.06  & \fs  0.03  & \fs -0.13   & \fs  0.06 &  \fs -0.16 & \fs  0.42 & \fs -0.00 & O3-V-O2 bending   & O3-V-O2 bending     \\
178 & 179 & 176 & \fs -0.02  & \fs  0.28  & \fs -0.40  & \fs  0.10   & \fs  0.04 &  \fs -0.22 & \fs -0.06 & \fs  0.18 & Na $\parallel$ c  & Na $\parallel$ c    \\
 90 & 90  & 111 & \fs -0.04  & \fs  0.30  & \fs  0.42  & \fs  0.13   & \fs  0.03 &  \fs -0.16 & \fs -0.08 & \fs  0.18 & chain rot.        & chain rot.          \\[2mm]
	\hline
   \end{tabular}
\caption{Calculated frequencies and eigenvectors of the A$_g$ phonon modes of \Na\ compared to experiment.}
\label{tab:eigenmodes_na}
\end{table*}
\begin{table*}
\vspace*{3mm}
\begin{tabular}{|cc @{\hspace{2mm}}c|cccccccc|cc|}
\hline
\multicolumn{3}{|c|}{Frequency (cm$^{-1}$)}  & \multicolumn{8}{c|}{Eigenvector} & \multicolumn{2}{c|}{Assignment}\\ 
\hline
\multicolumn{2}{|c}{ Experiment} 
 &Theory  & &&&&&&&&&\\[-.7mm]
\fs Ref.~\onlinecite{Konstantinovic00}&\fs Ref.~\onlinecite{Popovic02} & & \fs  V$_x$&  \fs V$_z$& \fs Ca$_z$& \fs O1$_z$& \fs O2$_x$& \fs O2$_z$&\fs O3$_x$& \fs O3$_z$ & Ref.~\onlinecite{Popovic02}& This work \\
	\hline
935 & 932   & 900 & \fs  0.06 & \fs  0.24 & \fs  0.01 & \fs  0.02 & \fs  0.02 & \fs -0.02 & \fs -0.08 & \fs -0.42 & V-O3 stretching   &  V-O3 stretching   \\
542 & 539   & 516 & \fs -0.01 & \fs -0.05 & \fs -0.01 & \fs  0.11 & \fs  0.49 & \fs  0.07 & \fs -0.02 & \fs  0.00 & V-O2 stretching   &V-O2 stretching   \\
472 & 470   & 446 & \fs  0.36 & \fs -0.16 & \fs  0.00 & \fs  0.39 & \fs -0.09 & \fs -0.05 & \fs -0.08 & \fs -0.00 & V-O1-V bending    & V-O1-V bending\\  
421 & 422   & 412 & \fs -0.34 & \fs -0.12 & \fs  0.02 & \fs  0.27 & \fs -0.05 & \fs -0.17 & \fs  0.18 & \fs -0.13 & O3-V-O2 bending   & O1$_z$+O3-V-O2 bending  \\
337 & ?	    & 307 & \fs  0.09 & \fs  0.20 & \fs  0.07 & \fs  0.10 & \fs -0.01 & \fs  0.20 & \fs  0.39 & \fs  0.05 & O3-V-O2 bending   &   O3-V-O2 bending\   \\
282 & 235.6 & 265 & \fs  0.19 & \fs -0.10 & \fs -0.14 & \fs -0.18 & \fs  0.09 & \fs -0.31 & \fs  0.25 & \fs -0.08 & O3-V-O2 bending   & O3-V-O2 bending\\
238 & 138.6 & 201 & \fs  0.02 & \fs  0.21 & \fs  0.39 & \fs  0.03 & \fs  0.06 & \fs -0.31 & \fs -0.05 & \fs  0.17 & Ca $\parallel$ c  & chain rot.\\
 91 & 90    & 106 & \fs -0.05 & \fs  0.32 & \fs -0.39 & \fs  0.15 & \fs  0.01 & \fs -0.15 & \fs -0.06 & \fs  0.19 & chain rot.	      & Ca $\parallel$  c 	\\[2mm]

	\hline
   \end{tabular}
\caption{Calculated frequencies and eigenvectors of the A$_g$ phonon modes of \Ca\ compared to experiment.} 
\label{tab:eigenmodes_ca}
\end{table*}

\begin{figure}[h!]
\begin{center}
\includegraphics[width=0.43\textwidth]{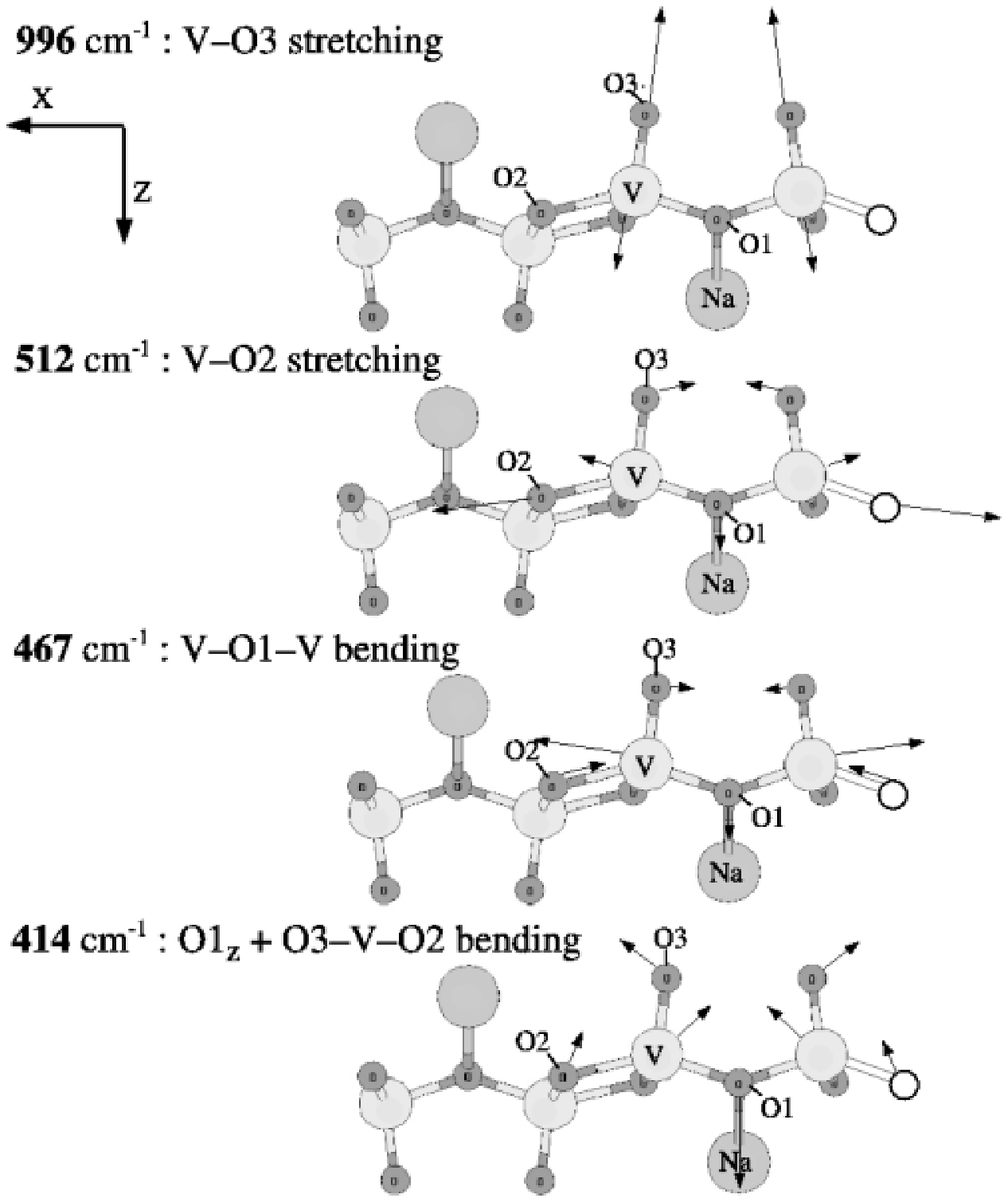}
\includegraphics[width=0.43\textwidth]{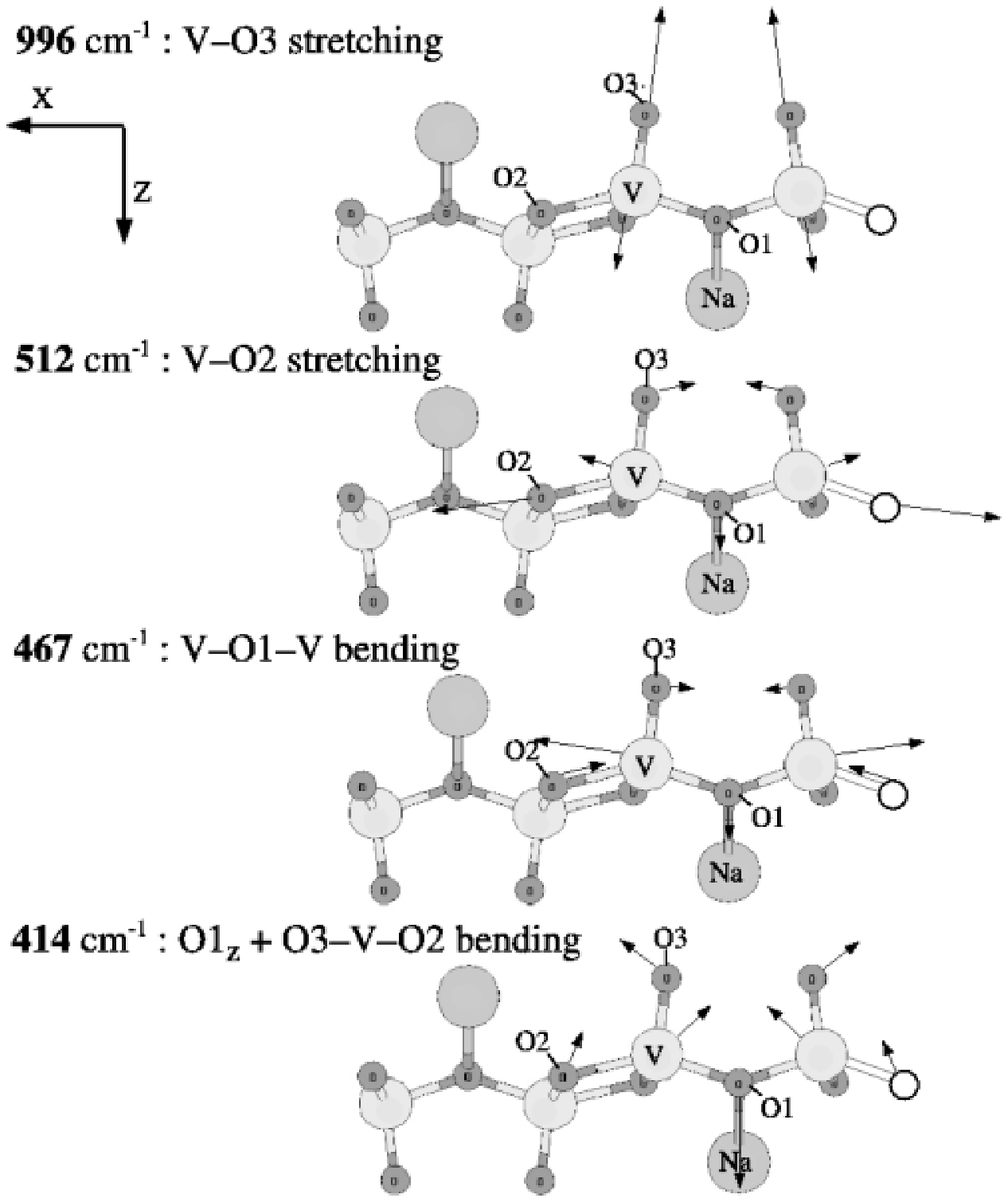}
\end{center} 
\caption{ Eigenvectors of \Na.}
\vspace{-5mm}
\label{fig:eigenvectors_na}
\end{figure}

For the calculation of the A$_g$ phonon modes we applied the {\it frozen-phonon} 
approximation.
To this extent, the atomic positions have been moved from their equilibrium. 
Four displacements (two in positive and negative direction, respectively) for each 
degree of freedom have been taken into account. The resulting forces were used to 
obtain the energy hyper-surface according to the procedure described 
in Ref.~\onlinecite{Ambrosch02} and to set up the dynamical matrix. 
Since the harmonic fully-symmetric ion vibrations do not change the occupancy of the V sites
from single (of fractional) to double, the electron
correlations do not significantly influence the elements of the dynamical
matrix.
At the same time, the correlation effects could be more important 
for the anharmonic terms relating the lattice forces and ion displacements.  

The frequencies of the  A$_g$ modes for 
\Na\ and \Ca\ are presented in Tables~\ref{tab:eigenmodes_na} 
and \ref{tab:eigenmodes_ca}, respectively, and compared to experimental data. 
The corresponding eigenvectors presented in these Tables, ${\bf e}_{\alpha\zeta}$,
are related to the real displacements
${\bf u}_{\zeta}^{\alpha}$ by
\begin{equation}
{\bf e}_{\alpha\zeta}={\rm const.}\times{\bf u}_{\zeta}^{\alpha}\sqrt{M_\alpha}, 
\end{equation} 
and are normalized as
\begin{equation}
\sum\limits_{\alpha=1}^{N}{\bf e}_{\alpha\zeta}^2=1,
\end{equation} 
where  $\alpha$ enumerates the ions with mass $M_\alpha$, $N=16$ is the number of ions
per unit cell, and $\zeta$ indicates the phonon mode.    
In order to visualize the lattice distortions according to the phonon eigenvectors, 
the corresponding atomic displacements  of \Na\ are shown in Fig.~\ref{fig:eigenvectors_na}, 
where the eigenvector components of Table \ref{tab:eigenmodes_na} refer to the 
equivalent positions labeled in the figure.
The phonon eigenvectors of \Ca\ are very similar, up to some differences discussed later.

For the eigenfrequencies of the \Na\ A$_g$ modes, good agreement with 
experiments\cite{Fischer99,Popovic99ssc,Popova99} is found, with deviations 
smaller than 5\%. Only for the lowest-energy mode the difference is 
larger. In full agreement with the results of Refs.~\onlinecite{Fischer99,Popovic99ssc} and \onlinecite{Popova99} 
the eigenvector of the 996~cm$^{-1}$ mode represents a stretching between V and the 
apical oxygen. Also the V-O2 stretching of the 512~cm$^{-1}$ mode and the V-O1-V bending of 
the 467~cm$^{-1}$ mode as suggested in Ref.~\onlinecite{Popovic99ssc} 
are verified, where we, however, find admixtures of an O1$_z$ 
motion for the former and an O2$_x$ movement for the latter, respectively. 
In the eigenvector of the 414~cm$^{-1}$ mode, the $z$-displacement of the in-rung oxygen
O1 is dominating, while this mode is described as pure O3-V-O2 bending 
in Ref.~\onlinecite{Popovic02}. At the same time, our theoretical frequencies are much 
closer to experiment (2\% deviation) than the calculated 
frequencies in Ref.~\onlinecite{Popovic02} (9\% difference). 
For the 308~cm$^{-1}$, the 232~cm$^{-1}$, the 176~cm$^{-1}$ and the 111~cm$^{-1}$  mode, 
the agreement of our results with the assignment of Popovic {\it et al.} 
\cite{Popovic02} is good. However, in most of the modes we find a more pronounced 
involvement of O1 compared to the interpretation of experimental results. 

For the eigenfrequencies of the \Ca\ A$_g$ modes, the agreement 
between theory and experiment\cite{Popovic02} is very good for the 
modes above 400~cm$^{-1}$. For these vibrations, the experimental 
assignment with respect to their symmetry is unambiguous.
The lower frequency of the apical oxygen vibration in \Ca\ compared to \Na\ is due to 
larger interionic distances, and hence, smaller force constants. 
A change of 5\% can be estimated within the Coulomb picture
from the different Vanadium charges in \Na\ and \Ca\  (i.e.~4.5 and 4, respectively), 
which is in qualitative agreement with experiment. 

The two lowest-frequency modes are swapped when Na is replaced by Ca, 
i.e.~the in-phase motion of Ca with the other atoms of the ladder (chain rotation)
  has higher energy than the out-of-phase vibration, where Ca vibrates~$\parallel$~c in opposite direction to the ladder.
  The frequency of the chain rotation mode (201~cm$^{-1}$) is roughly twice 
  as high as in \Na, representing a difficulty in the interpretation of the 
  measured modes of \Ca\ which in Refs.~\onlinecite{Popovic02} and \onlinecite{Konstantinovic00} was 
  done in comparison with the phonons of \Na.  In Ref.~\onlinecite{Popovic02} the 138~cm$^{-1}$ 
  mode of the unpolarized spectrum was interpreted as A$_g$ vibration since its frequency compared to 
  that of the (Na $\parallel$ c) mode scales as the inverse  square root of the corresponding masses.   
From our analysis we conclude, that the measured 138~cm$^{-1}$ mode is not an A$_g$ vibration. 
The physical origin of the frequency shift of the chain rotation mode is related to the 
stronger inter-ladder interaction in \Ca\, 
which is also reflected in the enhanced tight binding parameter $t_i$ (see Section IV B).
On the other hand, the higher Ca mass is mainly responsible for the decrease 
of the "Ca $\parallel$ c" mode frequency.
It is still unclear, however, whether the calculated 201~cm$^{-1}$ mode corresponds 
to the experimentally observed  213~cm$^{-1}$  vibration\cite{Popovic02} or, more probable, 
to a mode at 235~cm$^{-1}$ (Ref.~\onlinecite{Popovic02}) 
(or 238~cm$^{-1}$ according to Ref.~\onlinecite{Konstantinovic00}) 
which has been assigned as an A$_g$ mode in both papers. 
In the latter case our calculated frequency of 265~cm$^{-1}$ can be related to the measured 
282~cm$^{-1}$ vibration.\cite{Konstantinovic00}
A similar problem concerns the interpretation of one more A$_g$ mode somewhat above 300~cm$^{-1}$,
where no clear experimental assignment is available in Ref.~\onlinecite{Popovic02}. 
In this context, the interpretation of the Raman scattering intensities could be 
helpful for an unambiguous assignment as will be discussed in the next section.

\subsection{Electron-phonon and spin-phonon coupling}

When ions are shifted from their equilibrium positions, the changes of the
bandstructure are a measure for the electron-phonon interaction. 
Two types of coupling can be considered:
The first one is the Holstein coupling, where the site energies change
with the ion displacements, while the other one is due to changes of the hopping
parameters. 
At the same time, the exchange path, which is formed by transitions between different 
sites, is also influenced by the phonons. This effect results in spin-phonon coupling. 
To investigate the type of the electron-phonon coupling and its strength for each mode, 
in Tables~\ref{tab:parameters_na} and \ref{tab:parameters_ca} we display  the 
changes of various model parameters with the corresponding ion 
displacements ${\bf u}_{\zeta}^{\alpha}$ (up to $\sim0.05$~\AA). 
These are the hopping parameters $t_\parallel$, $t_\perp$, and $t_i$, the 
charge transfer gap $E_g$, and the exchange parameters $J_{\|}$ (along the ladders) 
and $J_{\perp}$ (within the rungs). $J_{\|}$ and $J_{\perp}$ can be estimated 
as $\sim t_{\|}^{2}/E_{g}$ and $\sim t_{\perp}^{2}/E_{g}$, respectively.\cite{Sherman99}
In this context we introduce the dimensionless phonon coordinate $Q$ by the relation
\begin{equation}
Q\sqrt{\frac{\hbar}{M_{\alpha}\omega_{\zeta}}}{\bf e}_{\alpha\zeta}=
{\bf u}_\zeta^\alpha.
\end{equation}

It turns out that only the higher-frequency phonons considerably modulate the one-ladder 
parameters $t_{\perp}$ and $t_{\parallel}$.  
The changes of the hopping matrix elements corresponding to $Q=1$ are below 0.025~eV, 
while they can be much larger in the energy shifts $\delta E_{g}$. Therefore the 
main mechanism of electron-phonon coupling can be assigned to Holstein-like interaction. 
Our results also allow to estimate the strength of spin-phonon coupling arising 
due to the phonon-induced modulation of the exchange parameters. The corresponding 
relative changes are summarized in Tables~\ref{tab:parameters_na} and \ref{tab:parameters_ca}.

\begin{table*}[hbt]
\begin{tabular}{|c|c|c|c|c|c|c|}
   \hline
$\omega_{\rm th}$ (cm$^{-1}$) & $\delta t_\|/\delta Q$ (eV) & $\delta t_\perp/\delta Q$ (eV) & $\delta t_{i}/\delta Q$ (eV)& $\delta E_{g}/\delta Q$ (eV) & 
($\delta J/J)_\|/\delta Q$  & ($\delta J/J)_{\perp}/\delta Q$  \\
\hline																					 
 996 & -0.0021    & -0.0099  &  0.0056 &	-0.0936 	 &  0.0190 &  0.0449\\
 512 & -0.0210    &  0.0213  & -0.0439 &	 0.0905 	 &  0.2952 & -0.0773\\
 467 & -0.0016    & -0.0115  &  0.0109 &	-0.0037 	 &  0.0052 &  0.0478\\
 414 & -0.0110    & -0.0069  & -0.0142 &	-0.0010 	 &  0.1284 &  0.0294\\
 308 &  0.0000    & -0.0167  & -0.0253 &	 0.0554 	 &  0.0003 &  0.0967\\
 232 & -0.0012    & -0.0012  &  0.0257 &	-0.0340 	 &  0.0085 &  0.0013\\
 176 &  0.0019    &  0.0061  &  0.0193 &	-0.0185 	 & -0.0130 & -0.0225\\
 111 &  0.0008    &  0.0028  &  0.0132&    -0.0117  	& -0.0055 & -0.0107\\
\hline
\hline
\multicolumn{7}{|c|}{undistorted:   $t_\|$ = 0.175 eV,  $t_\perp$ = 0.387 eV, $t_{i}$ = 0.117 eV, $E_{g}$ = 2.565  eV }    \\
\hline																							   
\end{tabular}
\caption{Parameters of electron-phonon and spin-phonon coupling for the A$_g$ eigenmodes of \Na. }
\label{tab:parameters_na} 
\end{table*}

\begin{table*}[hbt]
\begin{tabular}{|c|c|c|c|c|c|c|}
   \hline
$\omega_{\rm th}$ (cm$^{-1}$) & $\delta t_\|/\delta Q$ (eV) & $\delta t_\perp/\delta Q$ (eV) & $\delta t_{i}/\delta Q$ (eV) & $\delta E_{g}/\delta Q$ (eV) & ($\delta J/J)_\|/\delta Q$ & ($\delta J/J)_\perp/\delta Q$   \\
\hline									    										
900  &  -0.0029 & -0.0089 &  0.0078 & -0.1286 &  0.0316  &   0.0470\\
516  &  -0.0113 &  0.0120 & -0.0253 &  0.0503 &  0.1863  &  -0.0565\\
446  &  -0.0013 & -0.0232 &  0.0139 & -0.0228 &  0.0052  &   0.1399 \\
412  &   0.0056 & -0.0094 &  0.0223 & -0.0239 & -0.0815  &   0.0540\\
307  &   0.0016 & -0.0017 & -0.0035 &  0.0054 & -0.0141  &   0.0185 \\
265  &   0.0071 & -0.0050 & -0.0337 & -0.0048 & -0.0895  &   0.0285 \\
201  &  -0.0019 &  0.0035 &  0.0278 & -0.0080 &  0.0269  &  -0.0217 \\
106  &   0.0013 & -0.0055 & -0.0110 & -0.0009 & -0.0227  &   0.0317 \\
						    	  
   \hline
   \hline
 \multicolumn{7}{|c|}{undistorted:   $t_\|$ =0.143 eV, $t_\perp$ = 0.321~eV, $t_{i}$ = 0.244 eV, $E_{g}$ = 2.882 eV }    \\
\hline																							   
\end{tabular}
\caption{Parameters of electron-phonon and spin-phonon coupling for the  A$_g$ eigenmodes of \Ca. }
\label{tab:parameters_ca} 
\end{table*}

 The changes in the  matrix elements can be understood as a result of the altered interionic distances and the electronic on-site energies. For example, the large phonon-induced  decrease of $t_\|$ in the 512~cm$^{-1}$ vibration of \Na\ is a consequence of the larger V-O distance in the same leg (see Fig.~\ref{fig:eigenvectors_na}).
At the same time, the V-O distance along the $x$ axis changes from 0.274~\AA\ to 0.309~\AA\ at $Q=1$.
The  decrease of $t_\perp$ for the 308~cm$^{-1}$ mode is due to an enhanced $z$-axis distance between 
V and the in-rung oxygen O1 by 0.02 \AA\ going from the relaxed to the distorted structure with $Q=1$. 
As a consequence, also the energy difference of the V and O1 orbitals is increased, diminishing the hopping parameter $t_\perp$.
The very strong modulation of the inter-ladder hopping $t_{i}$ by the 512~cm$^{-1}$ mode is due 
to a zig-zag like deformation of the legs in the $(x,y)$ plane, i.e.~a vibration of neighboring vanadium and oxygen atoms of one leg in opposite direction,  
and hence an increase of this V-O$_2$ distance.

As can be seen in Tables~\ref{tab:parameters_na} and \ref{tab:parameters_ca}, the biggest 
change for both $t_\|$ and $t_\perp$, in \Na\ is caused by the V-O2 stretching mode, 
while in \Ca\ this mode leads to the most pronounced change only in $t_\|$, but the 
largest modulation of $t_\perp$ is due to the V-O1-V  mode (446~cm$^{-1}$). 
The reason for this can be found in the displacement of the in-rung oxygen O1 
along the $z$ axis, which is much larger compared to \Na.
For both compounds, the biggest effect on $E_g$ is observed for the V-O3 stretching mode.
 
We emphasize here that both, electron-phonon coupling 
(leading to a modulation of $E_g$ and the hopping matrix elements) and spin-phonon coupling 
(leading to a modulation of $J_{\|}$)
in \Na\ and \Ca, are considerably strong. 
\cite{Choi03}. For this reason the lattice distortion
in the low-temperature phase of \Na\ can be related to total-energy changes originating 
from charge as well as spin \cite{Suaud02,Hozoi02} degrees of freedom.  

\section{ Raman scattering}

With the knowledge of the phonon modes and the dielectric functions we can calculate the phonon Raman spectra of \Na\ and \Ca. For this purpose we use the approach developed in
Ref.~\onlinecite{Ambrosch02}, where at a given exciting light 
frequency $\omega_I$ the total Raman intensity $I_{R}$ at 
temperature $T$ in arbitrary units is:
\begin{equation} 
I_{R}(\omega_R)=\sum_{\zeta}(n_B(\omega_{\zeta})+1)
\left| 
\langle 1 |\frac{\partial \varepsilon_{ii}^{\omega_I}}{\partial Q} \hat{Q}| 0 \rangle \right|^2
L(\omega_R,\omega_{\zeta},\Gamma)
\end{equation}
Here $\omega_R$ is the Raman shift, the Cartesian indices $ii$ correspond to the polarizations
of incident and scattered light, which are the same due to the orthorhombic symmetry of the crystal.
$ |1\rangle$ and $| 0 \rangle$ denote the one-phonon and the phonon-less states, respectively, and $\hat{Q}$ is the operator of the phonon coordinate. $n_B(\omega_{\zeta})=1/(\exp{(\hbar\omega_{\zeta}/T})-1)$ is the 
phonon Bose distribution function, and $L(\omega_R,\omega_{\zeta},\Gamma)$ is the Lorentzian
shape of the phonon line with a broadening $\Gamma$, which was chosen to be 25~cm$^{-1}$ for all modes.

\begin{figure}[h]
\begin{center}
\includegraphics[width=0.4\textwidth]{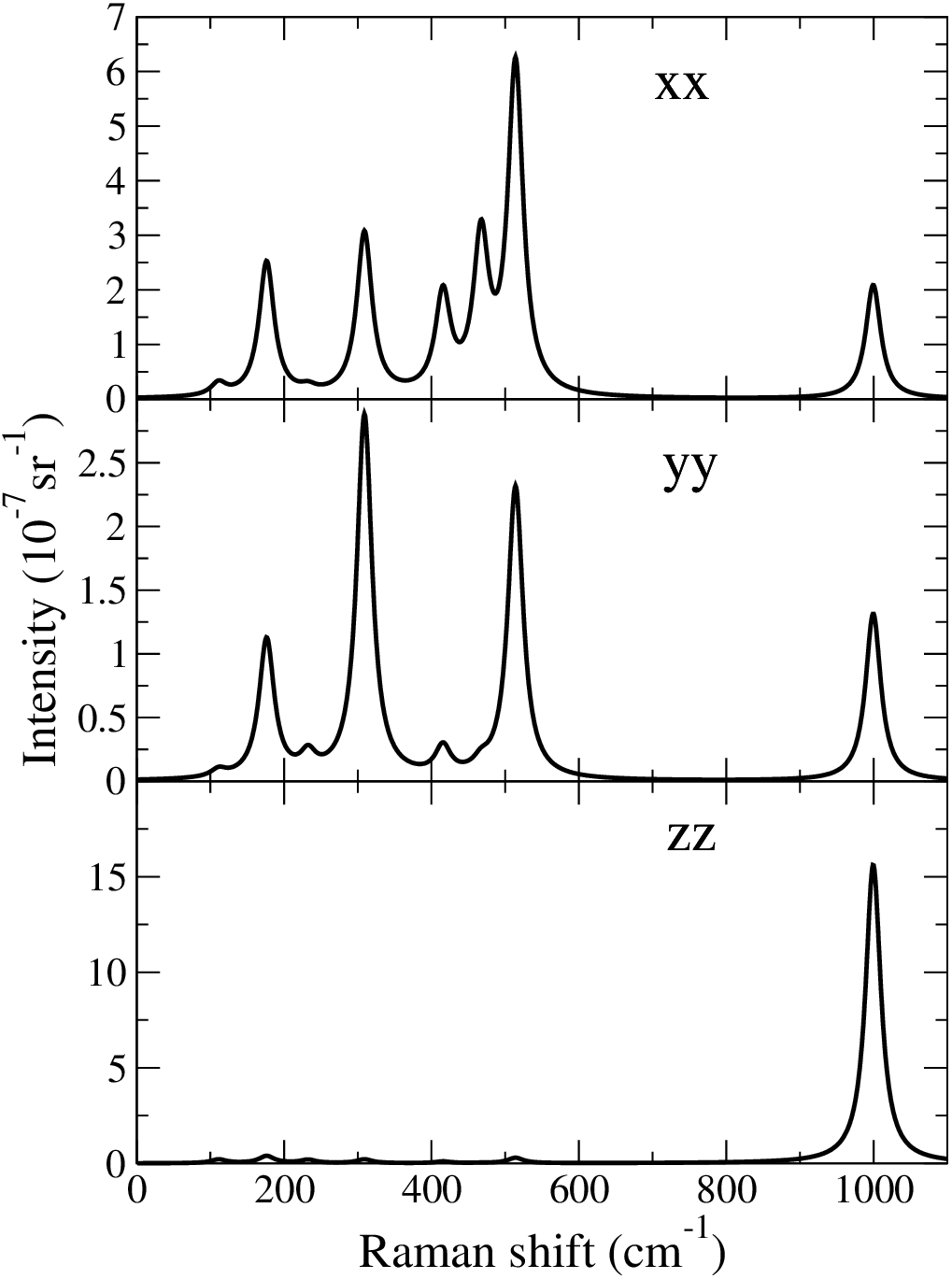}
\vspace{-2mm}
\end{center}
\caption{ Raman intensity ${I_R}$ for \Na\ in three different geometries
at $\omega=2.41$ eV ($\lambda=514.5$ nm) and $T=300$ K.}
\label{fig:raman_na}
\end{figure}

\begin{figure}[h]
\begin{center}
\includegraphics[width=0.4\textwidth]{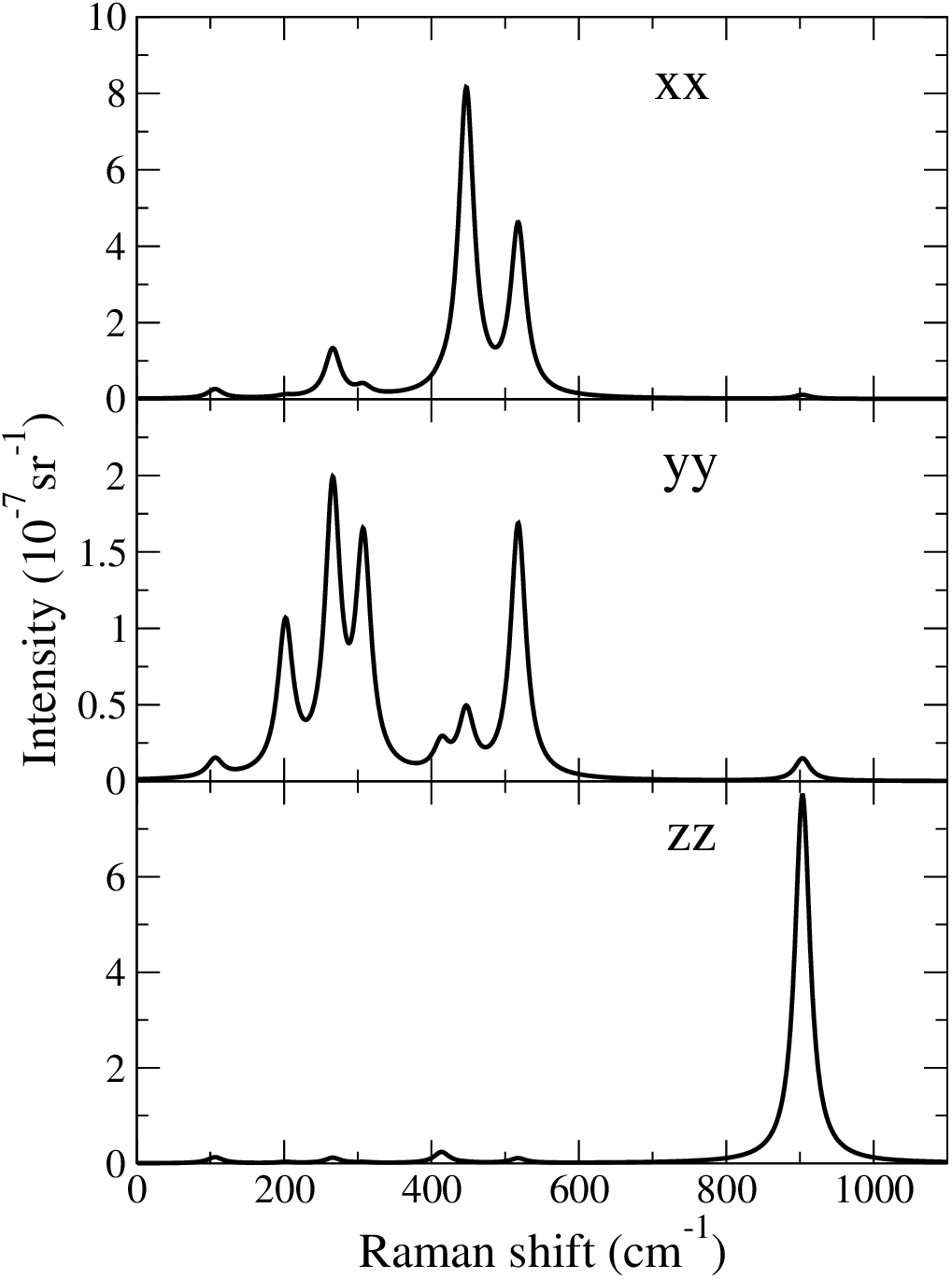}
\vspace{-2mm}
\end{center}
\caption{Raman intensity ${I_R}$ for \Ca\ in three different geometries
at $\omega=2.41$~eV ($\lambda=514.5$ nm) and $T=300$ K.}
\label{fig:raman_ca}
\end{figure}

The total Raman intensity, i.e.~the sum over all phonon contributions, is presented in Figs.~\ref{fig:raman_na} and \ref{fig:raman_ca} for \Na\ and \Ca, respectively, for an incident light energy of 2.41~eV ($\lambda=514.5$~nm), which is used in the Raman experiments available in the literature. In the $xx$ polarization seven out of eight modes are clearly visible, only the 232 cm$^{-1}$ vibration has negligible intensity, in excellent agreement with experiments.\cite{Fischer99,Popovic99ssc} We note that also the relative peak heights are fully reproduced. This scattering geometry exhibits the highest intensity for all modes except the highest one, which dominates the $zz$ polarized spectra. The intensity of all other modes in this polarization is two orders of magnitude smaller and hence hardly visible in the measured spectra. The only exception is the 110~cm$^{-1}$ mode, where theory cannot reproduce the experimentally observed sharp peak. One possible explanation could be provided by the extremely pronounced resonance behavior of most of the vibrations as will be discussed below. The $yy$ polarization exhibits intensities in between the magnitudes of the $xx$ and $zz$ counterparts. 
 In accordance with experiment the 467~cm$^{-1}$ vibration is absent in this scattering geometry.

For \Ca\ the situation is similar. The highest energy mode exhibits the highest intensity in $zz$ polarization. Between 400 and 600~cm$^{-1}$ the $xx$ intensities are dominating. Only below 400~cm$^{-1}$, the $yy$ spectra are comparable in magnitude or even bigger.

\begin{figure}[h!]
\begin{center}
\includegraphics[width=0.42\textwidth]{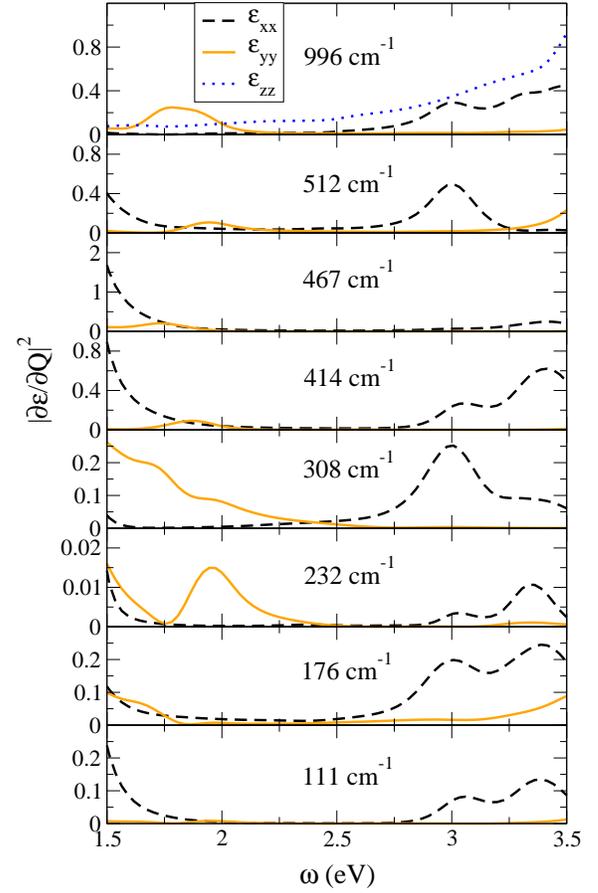}
\vspace{-2mm}
\end{center}
\caption{ Derivatives of the dielectric functions $|{\partial\varepsilon}/{\partial Q}|^2$ with respect to 
displacements along the eigenvectors of the phonon modes of \Na\ as indicated by their phonon frequencies. 
The $zz$ component is considerable only for the highest frequency  and is therefore omitted in the other panels.}
\label{fig:dEpsilon_na}
\end{figure}

The Raman scattering intensity is governed by the change of the crystal polarizability with the nuclei vibrating around their equilibrium positions.
The dependence of the dielectric function on the phonon coordinate arises due to two main reasons, i.e.~the $Q$ dependence of the momentum matrix 
element and the $Q$-dependence of the interband transition energy.\cite{Sherman03}
The latter contribution leads to a stronger resonance behavior of the Raman
intensity than the former. 
To illustrate  the influence of the lattice 
vibrations on the dielectric function, in Fig.~\ref{fig:dEpsilon_na} we present 
 $|\partial\varepsilon_{ii}(\omega)/{\partial Q}|^2$
for the phonon modes  of \Na. 
In the energy range  $\omega>2.5$ eV the derivative of the $xx$ component 
dominates over $|\partial\varepsilon_{yy}(\omega)/{\partial Q}|^2$ for all modes. 
In the low energy region the latter shows a dramatic resonance behavior 
for the 232~cm$^{-1}$ and the 308~cm$^{-1}$ vibration, and is comparable or 
even slightly larger in magnitude than $|\partial\varepsilon_{xx}(\omega)/{\partial Q}|^2$ around 2~eV for 
the next three modes higher in energy. All modes but the highest one exhibit extremely 
strong resonances in the infrared in either of the two polarizations. For the highest mode, 
the $zz$ polarization is dominating, while only below 2~eV the $yy$ component becomes 
more pronounced. We thus predict much higher Raman scattering intensities for incident 
light frequencies both higher and lower than those used in experiments so far. 
We mention in this context, that Fischer {\it et al.}\cite{Fischer99} observed strong 
relative intensity changes 
in $xx$ geometry when measuring the low temperature phase using different photon energies.

A similarly strong resonance behaviour is also expected for \Ca. While with the 
exception of the highest mode the intensities of the $zz$ component are small 
and the influence of the photon energy is negligible in \Na, this scattering 
geometry is slightly richer for \Ca, with generally slightly higher relative intensities 
and an even more pronounced resonant behavior of the highest mode.

\section{Conclusions}

In this paper, we have studied optical properties and lattice dynamics
of  \Na\  and \Ca\ in the $Pmmn$ phase from first principles. The calculations are based on the theoretically optimized crystal structures obtained within the generalized gradient approximation. 
Effective bandstructure parameters have been extracted by mapping our 
results onto a tight-binding model. We have obtained 
the Hubbard repulsion $U$ on the V sites being approximately 2.45 eV, both for
\Na\ and \Ca. The dielectric functions have been determined within the random phase approximation and are in very good agreement with available experiments. Our results show that the 1~eV
peak in the $xx$ component arises due to transitions between
the bonding and the antibonding combination of V~$d_{xy}$ orbitals within one rung.    
By diagonalizing the dynamical matrix we have obtained the phonon frequencies
for the fully symmetric vibrations which are in very good agreement with measured data.  
With the knowledge of the phonon eigenvectors and the changes of the bandstructure
caused by the phonon modes we have estimated the parameters of electron-phonon 
and spin-phonon coupling for both compounds. We find that the strongest contribution to the electron-phonon coupling comes from the phonon-modulation of the charge transfer gap $E_g$.
At the same time, other effects arising from altered hopping matrix elements
can be important. Finally, we have calculated the phonon Raman spectra of these compounds and analyzed the frequency-dependent dielectric function modulated by the ion displacements according to the lattice vibrations.  
On this basis we predict a strong resonance behavior for both \Na\ and \Ca.

As an outlook for further investigations, this detailed analysis provides a basis for comparison with the low-temperature phase. Moreover, the Hubbard parameters extracted in this work can be used as input for further calculations by, e.g., Exact Diagonalization\cite{Aichhorn03,Edegger} or Quantum Monte Carlo simulations\cite{Gabriel03}

\vspace{3mm}
\noindent{\bf Acknowledgment}  The work is financed by the Austrian Science  
Fund (FWF), project P15520. We also appreciate discussions with
M. Aichhorn and support by the FWF project 
P16227 and the EU RTN network EXCITING (contract HCPR-CT-2002-00317). 
EYS is grateful to R.~T.~Clay, A.~Damascelli, P.~Lemmens, S.~Mazumdar, and M.~N. Popova 
for interesting discussions and suggestions.

\end{document}